\documentstyle[aps,epsfig,float,twocolumn,prl]{revtex}
\newcommand{\be}{\begin{equation}}
\newcommand{\ee}{\end{equation}}

\begin{document}
\twocolumn[\hsize\textwidth\columnwidth\hsize\csname @twocolumnfalse\endcsname
\draft
\title{Spin-Orbit Scattering and Time-Reversal Symmetry: Detection
of a Spin by Tunneling}
\author{M. B. Hastings}
\address{
Center for Nonlinear Studies and Theoretical Division, Los Alamos National
Laboratory, Los Alamos, NM 87545, hastings@cnls.lanl.gov
}
\date{January 6, 2004}
\maketitle
\begin{abstract}
We consider the possibility of detecting spin precession in a magnetic
field by nonequilibrium transport processes.  We find that time reversal
symmetry imposes strong constraints on the problem.  Suppose
the tunneling occurs directly between systems at two different chemical
potentials, rather than sequentially
via a third system at an intermediate chemical potential.  Then,
unless the
magnetic fields are extremely strong or spin polarized electrons are used,
the periodic signal in the current results from beating together two different
precession frequencies, so that observing a signal near the Larmor frequency
in this case requires having some cluster with a $g$ factor close to zero.
\vskip2mm
\end{abstract}
]

Transport of spin-polarized current through
an STM\cite{stm,avb} opens the possibility of investigating magnetic systems
at a much smaller scale than is possible using electron spin 
resonance\cite{esr}.  The ability to manipulate single
spins via STM is a key part of the development of spintronics\cite{sp} and
quantum information processing.

Interestingly, even using spin-unpolarized electrons, a periodic modulation
of tunneling current through an STM tip was observed experimentally\cite{s}.
The original proposed explanation depended on periodic modulation of charge
at the Larmor frequency by a precessing spin center.  Another
very interesting possibility depends on spin-orbit scattering\cite{d,wt32,wt33}.
However, in this
paper, we will show that, with two exceptions,
time reversal symmetry imposes strong constraints
on the problem, and for both of the mechanisms considered leads to
modulation at a frequency resulting from beating together two different
precession frequencies.  Then, for a single impurity spin, the
modulation in the current appears at
twice the Larmor frequency, and not at the Larmor 
frequency itself.  To obtain a signal at the Larmor frequency requires
two spins, one of which has a vanishing $g$-factor\cite{av}.

The first possible exception is to have
a sufficiently large magnetic field to introduce Aharonov-Bohm phases
for the electrons; in this case time-reversal symmetry of the tunneling
Hamiltonian is broken by the external field.
This case is unlikely to be experimentally relevant, 
due to the strength of fields required. 
The more interesting exception
involves sequential tunneling through a quantum dot\cite{d,dm},
and will be discussed more below.  Sequential tunneling refers
to a system in which there are two leads and a quantum dot, with the
chemical potential one lead far above that of on the dot,
and the chemical potential on the other lead far below that of the dot.
Thus, this system has three different chemical potentials;
the results in this paper will apply to systems in which there are only two
different chemical potentials.

As a model, we first
consider a tunnel junction with a single nearby spin.  The electrons
can couple to the spin via exchange coupling, and there
may also be a spin-orbit coupling present.  This
provides a means of measuring the spin in terms of its effect
on the tunnel current.  We first consider
the appropriate Hamiltonian for the system in the absence of any
external magnetic field, so that the system is time reversal invariant,
and then consider the effect of any applied magnetic field.  We find
that the effect of time reversal symmetry strongly limits the allowed
modulation of the current by the spin.  Our physical setup
with an impurity spin and a single tunnel junction will
be similar to that considered in \cite{wt3}; similar results with some 
exceptions
hold for tunneling through a quantum dot as considered in \cite{d}, as discussed
below.

{\it Time Reversal Symmetry---}
Consider the most general possible Hamiltonian, written as
${\cal H}={\cal H}_L+{\cal H}_R+{\cal H}_T$, where ${\cal H}_{L,R}$ are the
Hamiltonians for the left and right leads of the system and ${\cal H}_T$
includes the tunneling terms, as well as terms coupling the electrons
to the spin.  The two leads
may be taken to be Luttinger liquids.  There is a voltage
difference $V$ between the leads, giving a chemical potential
difference $\mu=qV$, where $q$ is the electron charge.  
We assume that the effects of
electron interaction are negligible in the tunneling region, as the
tunneling process is weak and only one electron tunnels at a time.  Thus,
we restrict to terms which are at most second order in the Fermion operators
in ${\cal H}_T$.  

Then, given the requirement of time reversal
symmetry, the most general possible ${\cal H}_T$, is
\begin{eqnarray}
\label{genh}
{\cal H}_T=t_0 T_0+t_1 J_0^j v_j
+t_2 J_0 (S_j w_j)\\ \nonumber
+J_{\beta\gamma}(\psi_{\beta}^{\mu})^{\dagger}S^{\mu\nu}
\psi_{\gamma}^{\nu},
\end{eqnarray}
where a sum over repeated indices is assumed.
Here $S_j$ (for $j=x,y,z$) is the spin operator on the impurity spin (which is
taken to have total spin $S$, where $S$ may be larger than $1/2$) and
$S^{\mu\nu}=\sigma_j^{\mu\nu} S_j$, where
$\sigma_j$ are the
Pauli spin matrices.  
The letters $\beta,\gamma$ identify the particular
lead: $\beta,\gamma=L,R$, while
$\psi^{\mu}$ are electron destruction operators with spin 
$\mu=\uparrow,\downarrow$.
We have defined the following operators:
first, a tunneling operator $T_0\equiv 
(\psi_{L}^{\mu})^{\dagger}\psi_{R}^{\mu}+h.c$.  Next, an electron spin
current
$J_0^j \equiv i\sigma_j^{\mu\nu}
[(\psi_{L}^{\mu})^{\dagger}\psi_{R}^{\nu}-h.c.]$. 
Finally, an operator
related to the current operator,
$J_0\equiv i[(\psi_{L}^{\mu})^{\dagger}\psi_{R}^{\mu}-h.c.]$.
The term in $t_0$ is a tunneling term.  
The term in $t_1$, which
combines spin-flip with tunneling, can arise via a mechanism
similar to that generating spin scattering by nonmagnetic impurities 
in semiconductors\cite{sf}, as noted in \cite{d}.
The fixed vector $v_j$ is determined by the spin-orbit
scattering.  The 
4 couplings $J_{\beta\gamma}$ are exchange couplings, 
while the fixed vector $w_j$ is set by 
both spin-orbit and exchange interactions.

The terms in (\ref{genh}) are time reversal symmetric\cite{trs}.
This operation complex
conjugates all terms in the Hamiltonian, and sends $\psi^{\uparrow}\rightarrow
\psi^{\downarrow}$ and $\psi^{\downarrow}\rightarrow-\psi^{\uparrow}$,
and similarly for $\psi^{\dagger}$.  Further, it changes the sign of
the spin operator, $S_i\rightarrow -S_i$.
Then, $J_0$ changes sign under time reversal as does $S$,
while $J_0^j$ is time reversal invariant.

The coupling of the tunneling current to the spin depends on a combination
of
exchange interaction between the electrons and spin with spin-orbit
scattering.  Since both of these interactions are in the term with $t_2$,
we focus on the following Hamiltonian:
\be
\label{specific}
{\cal H}_e={\cal H}_L+{\cal H}_R+t_0 T_0^{\mu\mu}
+t_2 J_0 S_z
\ee
Here, we have
chosen $w$ to lie along the $z$-direction.  However, many of the results
we obtain will be valid for much more general Hamiltonians, as will be
discussed at the end.

{\it External Magnetic Fields---}
To Eq.~(\ref{specific}),
we add an external magnetic field.  This field breaks the
time reversal symmetry and allows for terms which are
odd under time reversal, so that ${\cal H}={\cal H}_e+{\cal H}_o$.  
Two particularly important allowed terms are
\be
\label{trb}
{\cal H}_o=B_i S_i + t_3 J_0.
\ee
The first term is the coupling of the field to the spin, while the
second results from a possible Aharonov-Bohm phase on the electrons traversing
the tunnel junction: the phase of the term coupling to $S_z$ has been changed
by the magnetic field, so that it is no longer imaginary.  We expect
that terms in $t_3$
can be disregarded for experimentally relevant fields.
 
The possible introduction of terms like $t_3$ is
very important.  To see this importance, return to the Hamiltonian
${\cal H}_e$.  In this Hamiltonian, the only spin operator that
appears is the operator $S_z$.  Thus, if we pick a basis for $S$ that
diagonalizes $S_z$ we can consider $S$ to be constant in time, and
thus $S_z$ in the Hamiltonian (\ref{specific}) can be considered as
a classical parameter.  Then, this Hamiltonian is equal to
${\cal H}_L+{\cal H}_R+
[z (\psi_{L}^{\mu})^{\dagger} \psi_{R}^{\mu}+h.c.]$,
where $z=t_0 + i t_2 S_z$.  Since the phase of $z$ is unimportant,
physical parameters like the current can depend only on the modulus:
$|z|=|t_0 + i t_2 S_z|=t_0^2+t_2^2 S_z^2$.

Thus, the current depends on $S_z^2$, but cannot depend on the
sign of $S_z$.  In fact, for the most general Hamiltonian of Eq.~(\ref{genh}),
this remains true, as will be discussed more below.
On the contrary, if we consider the combined Hamiltonian,
${\cal H}_e+t_3 J_0$, the 
modulus becomes $t_0^2+t_3^2 + 2 t_2 t_3 S_z + t_2^2 S_z^2$,
and thus does depend on the sign of $S_z$.  This difference in the
dependence of the current on $S_z$, when $S_z$
is considered as a classical field (unchanging in time)
determines, as we will see, how strong
the modulation of the current is at the Larmor frequency, $\omega_L$, when
$S_z$ acquires time dynamics due to the introduction of an external magnetic
field.
The Hamiltonian ${\cal H}_e+{\cal H}_o$, with $t_3\neq 0$,
has been considered previously
in the context of a two level system rather than a spin, in which case such
terms do not violate time reversal symmetry\cite{wt3}, as well as in
the context of a spin system for which such terms do violate time
reversal symmetry\cite{wt32,wt33}.  These
authors found a modulation in the tunneling current at the Larmor
frequency.  In the case of tunneling through a dot, as considered in \cite{d},
similar time reversal symmetry violating terms were considered in the
matrix $\Omega_{rss'}$ of that paper, again finding modulation at the
Larmor frequency.

{\it Current Modulation at Weak Field---}
We now consider the case instead with a sufficiently weak field, so that
$t_3$ can be disregarded:
${\cal H}={\cal H}_e+B_i S_i$.
We will find only a very weak
modulation of the current at the Larmor frequency, and a much more
substantial modulation at twice the Larmor frequency.  This has
important experimental implications for the detection of a precessing
spin with a tunneling current: for small $t_3$, a much stronger signal
should be found at twice the Larmor frequency in this setup with
a single tunnel junction.

The effective dynamics of the density matrix for the spin can be
found using previous results\cite{effective}, based on a
Keldysh technique\cite{k}.  Throughout, we
work perturbatively in the tunneling, ${\cal H}_T$.
Define $I(\mu)=2\pi\rho_L\rho_R
\mu^{\alpha+1}/\Gamma(\alpha+2)$.
The parameter $\alpha$ depends on the
Luttinger parameters of the leads and the sample geometry; it is equal to
zero for Fermi liquid leads\cite{pa}.
The product of the density of states in the
left and right leads at energy $\mu$ is $\rho_L\rho_R \mu^{\alpha}$.
Then, the result for the effective dynamics is
\be 
\label{diss} 
\dot\rho=-i\Bigl[{\cal H}_0,\rho\Bigr]+
\frac{t_2}{2} 
\frac{{\rm d}I}{{\rm d}\mu}\Bigl[S_z,\Bigl\{\Lambda,\rho\Bigr\}\Bigr]-
t_2^2\frac{I(\mu)}{2}\Bigl[S_z,\Bigl[S_z,\rho\Bigr]\Bigr], 
\ee
where $\Lambda=[{\cal H}_0,t_2 S_z]$, and ${\cal H}_0$ is
the Hamiltonian for the spin.  This includes the external field,
$B S_x$, as well as an additional effective field along the $z$
direction due to the coupling to the electron current.  We choose to
absorb the additional field into a renormalization of the bare field, so
that the net field is along the $x$-direction: ${\cal H}_0=B S_x$.
Thus, $\Lambda=-i t_2 B S_y$.

The density matrix evolution (\ref{diss}) describes the Hamiltonian
evolution of the spin under ${\cal H}_0$, as well as dissipation (the
second term) and decoherence (the third term).
For large $V$, Eq.~(\ref{diss}) leads to an effective temperature
$T_{\rm eff}=(\alpha+1)^{-1}qV/2$.  This gives a stationary distribution
of the density matrix for the spin:
\be
\rho=\frac{1}{2S+1}(1-BS_x/T_{\rm eff}).
\ee

{\it Current---}
We now consider the average current across the system.  The current
operator $J$ is defined by taking a derivative of the Hamiltonian
with respect to an external gauge field, giving
\be
\label{hdef}
J=q(t_0 J_0-t_2 T_0 S_z).
\ee
The average
current is, for $B<<\mu$, \cite{effective}
\begin{eqnarray}
\langle J \rangle=q \langle t_0^2 + t_2^2 S_z^2 \rangle 
I(\mu)
-qB t_2^2
\langle [S_z,S_x]S_z\rangle
\frac{{\rm d} I(\mu)}{{\rm d} \mu}.
\end{eqnarray}
The first term is the current that results from considering $S_z$ to be
a classical object, while the second term is a weak quantum correction
that slightly reduces the current due to fluctuations in $S_z$.

As long as the net magnetic field is along the $x$-direction, 
$\langle S_z\rangle$=0, and $\langle S_z^2\rangle=(1/3)S(S+1)$.
If the field acquires a $z$-axis component, then, for $S>1/2$, the
expectation value $\langle S_z^2 \rangle $ will increase, and thus
the net current will increase.  The expectation value
$\langle [S_z,S_x]S_z \rangle=(B/T_{\rm eff})(1/6)S(S+1)$.
Thus, for a net field along the $x$-direction, the net current is
\begin{eqnarray}
\langle J \rangle=q (t_0^2+(1/3)S(S+1)) I(\mu)
\\ \nonumber
-q t_2^2 (B^2/T_{\rm eff}) 
(1/6)S(S+1)
\frac{{\rm d}I}{{\rm d}\mu}.
\end{eqnarray}

{\it Current Noise---}
We now wish to compute the current-current correlator,
$\langle J(\omega) J(-\omega) \rangle$.
At second order in ${\cal H}_T$, we obtain the shot
noise contribution.
It is obtained by using Eq.~(\ref{hdef}) for $J$ and computing the
correlation function
$\langle J(\omega) J(-\omega) \rangle$ using the Hamiltonian 
${\cal H}_L+{\cal H}_R$,
and no higher powers of ${\cal H}_T$.
The result is
$\langle J(\omega) J(-\omega)\rangle=2 q \langle J \rangle$, for
$\omega,\omega_L<<\mu$.

At fourth order in ${\cal H}_T$, many contributions are possible.  The
full calculation is quite involved, however the most important
contribution is a ``modulation" of the current.  We consider a modulation
of the current to be a term in $\langle J(\omega) J(-\omega) \rangle$ which
is of order $\mu^{2(\alpha+1)}$.  In contrast, for example, the
shot noise is only of order $\mu^{\alpha+1}$.

Such modulation terms are obtained as follows\cite{effective}:
we first consider $S_z$ to be a classical field, and obtain the
dependence of the current on $S_z$, giving
$J=q (t_0^2+t_2^2 S_z^2) I(\mu)$.
Then, we consider how $S_z$ varies in time, using the
dynamics (\ref{diss}), to compute the correlation functions of $S_z$.
The result is
\be
\langle J(0)J(t)\rangle=q^2 t_2^4 
I(\mu)^2 \langle S_z^2(0)S_z^2(t)\rangle.
\ee
For $S>1/2$, this gives a modulation of the current at twice the
Larmor frequency.  However, there is no modulation at the Larmor frequency
itself for any $S$.

Diagrammatically, we show a term giving such modulation in Fig.~1(a).  
Lines with arrows represent electron propagators, while pairs of straight lines
without arrows represent spin propagators 
(here, simply to give some diagrammatic way of writing
interactions with spins, we have chosen to use the standard represention of
a spin operator as a pair of fermion operators).  The curly lines
the current operator, while the crosses represent
the operator ${\cal H}_T$; since these two operators have straight lines
leaving them, this indicates that we take the term
$-qt_2 T_0 S_z$ in the current operator, and the term 
$t_2 J_0 S_z$ in ${\cal H}_T$.  There are two pairs of
lines connecting the bubbles,
giving a modulation at twice the Larmor frequency.  In Fig.~1(b), we
show a similar diagram with only one pairs of lines connecting the bubbles;
however, this diagram vanishes.

The fact that the modulation is not present at frequency $\omega_L$
will be true to all orders.  In contrast, other terms in 
$\langle J(-\omega)J(\omega)\rangle$, which are lower order in $\mu$ and
thus do not represent a modulation for this
experimental setup, may show a peak at the Larmor frequency.  
These terms all depend on $S_z$ having a time dependence on time
scales of order ${\mu}^{-1}$ (the time over which a tunneling event
occurs), and thus all of these terms will be suppressed by powers of $B/\mu$ and
are therefore expected to be quite small, for this situation.  In
such terms, we would find diagrams with both current operators
on the same electron bubble, rather than on different bubbles as
was the case in Fig.~1.

However, for other situations, these terms may be important.
Consider sequential
tunneling from one lead to the other via a quantum dot\cite{d},
where the chemical potential in one lead is far higher in energy than the dot,
and the chemical potential in the other lead is far lower in energy than the
dot.  In this case, the current through the system is largely independent
of the exact difference, $\mu$, between the chemical potentials in
the two leads.  Our considerations in this section relied on computing
the order of terms in $\mu$.  However, in the given experimental setup with
the quantum dot, the current is of order $\mu^0$ and the time to tunnel
from left to right lead is determined not by $\mu$, but
by the hopping matrix elements between
the quantum dot and the leads.
Then, it has been shown that\cite{dm} for tunneling through a quantum
dot in this sequential tunneling
regime, one can obtain a contribution to the current
at the Larmor frequency which is comparable to the zero frequency component
of the current.  Thus, while in \cite{d} terms were considered which
violate time reversal symmetry, with slight modifications a signal at
the Larmor frequency can be found using a Hamiltonian which respects
time reversal symmetry.

{\it More General Hamiltonians--}
The reason for a lack of modulation is that if the impurity spin is
held fixed, with a static $S_z$, then the current depends only on $S_z^2$ and 
not on $S_z$ itself.   For the Hamiltonians considered above,
this results from the fact that the current depends only on the modulus of
the hopping coefficient $z$ above.
However, this is true even in a more general case.
Consider any Hamiltonian including potential, interaction, and
position-dependent coupling of the electron current to $S_z$:
${\cal H}=\psi^{\dagger}(x)(\partial^2/2m+U(x))\psi(x)+
\psi^{\dagger}(x)\psi(x)\psi^{\dagger}(y)\psi(y)V(x-y)+
i\psi^{\dagger}(x)\{F(x),\partial_x\}\psi(x)S_z$.
The spin-orbit coupling introduces an effective vector potential.  If
this vector potential has a non-vanishing curl, then it leads to a non-vanishing
effective
magnetic field and it {\it is} possible for the current to depend on the sign
of $S_z$.  However, in cases like those considered above, in which this
effective
vector potential acts only across a single tunnel junction and does
not produce any magnetic field, the current does not depend on the sign
of $S_z$, only on its magnitude.  Further, in many cases with non-vanishing
effective field, the current will still depend only on the sign of $S_z$,
as must be determined on a case-by-case basis; for many systems, the
magnetoresistance curve is symmetric in the magnetization, for 
example\cite{tbp}.

{\it Example with Modulation---}
Let us finally give an example that does leads to a 
modulation at the Larmor frequency.  Suppose the Hamiltonian is
${\cal H}_L+{\cal H}_R + {\cal H}_T+B S_x$, with
${\cal H}_T=t_0 T_0
+t_2^a J_0 (S_j^a w_j)
+t_2^b J_0 (S_j^b w_j)$ and
${\cal H}_o=
g^a B_i^a S_i^a+
g^b B_i^b S_i^b$, where there are two impurity spins, $a,b$, with different
$g$-factors, coupled to the current.
Then, choosing $w$ to lie along the $z$-direction, the
current modulation is \be
\langle J(0)J(t)\rangle=I(\mu)^2
\langle \Omega(0)^2\Omega(t)^2\rangle,
\ee
where we define $\Omega=t_2^a S_z^a + t_2^b S_z^b$.  Then, if the
two spins have different precession frequencies $\omega_L^{1,2}$,
the two spins will ``beat" together and
peaks will be observed at $\pm 2\omega_L^a,\pm 2\omega_L^b,
\pm\omega_L^a\pm\omega_L^b$.
If one spin precesses slowly, $\omega_L^b\approx 0$, then this gives
a peak at $\omega_L^a$.
Alternately, the same result can be obtained by tunneling spin-polarized
electrons, again giving rise to a modulation at $\omega_L$.

{\it Conclusion---}
In conclusion, we have shown that the assumption of time reversal
symmetry severely constrains the ability to measure a spin via
a tunneling current.  For weak magnetic fields, where we can neglect
possible Aharonov-Bohm phases, the modulation is observed at twice
the Larmor frequency, rather than at the Larmor frequency.  This result
is fairly general for systems with a single tunnel junction; for systems
with a quantum dot in the sequential tunneling regime, or other systems
in which the tunneling current is largely independent of the voltage,
a peak in the current-current correlator at the Larmor frequency is possible.
While we
have derived this result for a specific Hamiltonian (\ref{specific}), it
holds for a more general Hamiltonian, if we take ${\cal H}={\cal H}_L+
{\cal H}_R+{\cal H}_T+B S_x$, with ${\cal H}_T$ given by Eq.~(\ref{genh}).

Given that such a modulation is observed at the Larmor frequency, and
assuming the Aharonov-Bohm phases are negligible,
then either the system is in a sequential tunneling regime or
else there are multiple spins present, enabling a beating of the precession
frequencies of the two spins.  However, in the latter case there should be a
current modulation at twice the Larmor frequency, at least for impurity
spins greater than $1/2$,
and thus an experimental measurement of that peak is suggested.

{Acknowledgments---}
I thank D. Mozyrsky, I. Martin, and A. V. Balatsky for useful
discussions.  This work was supported by US DOE W-7405-ENG-36.

\begin{figure}[!t]
\begin{center}
\leavevmode
\epsfig{figure=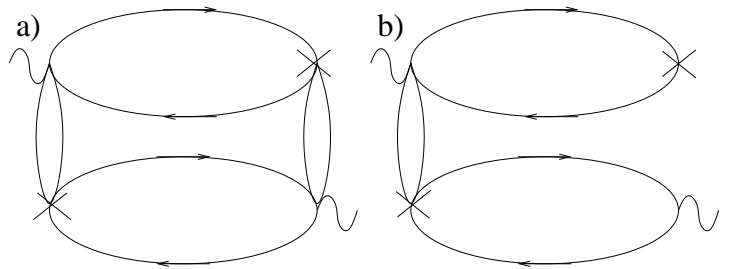,height=5cm,angle=0,scale=.7}
\end{center}
\caption{(a) Example diagram showing modulation at $2\omega_L$. (b) Diagram
which vanishes.}
\end{figure}
\end{document}